\documentclass[preprint]{raa}            

\usepackage{graphicx,times}             
\usepackage{natbib}
\usepackage{amssymb,amsmath}
\bibpunct{(}{)}{;}{a}{}{,}

\usepackage[pagebackref=true]{hyperref}
\hypersetup{colorlinks = true, linkcolor = green, anchorcolor = red, citecolor = blue, filecolor = red, pagecolor = red, urlcolor = red}

\newcommand{\mpch}{{\rm Mpc}\ h^{-1}}
\newcommand{\hmpc}{h\ {\rm Mpc}^{-1}}
\newcommand{\ba}{\begin{eqnarray}}
\newcommand{\ea}{\end{eqnarray}}

\begin{document}

\title{Expansion series of the pairwise velocity generating function and its implications on redshift space distortion modeling
}

   \volnopage{Vol.0 (20xx) No.0, 000--000}      
   \setcounter{page}{1}          

    \author{Junde Chen
      \inst{1}
    \and Pengjie Zhang
      \inst{1,2,3}
    \and Yi Zheng
      \inst{4}
   }

    \institute{Department of Astronomy, Shanghai Jiao Tong University, Shanghai 200240, China {\it jundechen@sjtu.edu.cn}\\
        \and
            Shanghai Key Laboratory for Particle Physics and Cosmology, People's Republic of China\\
        \and
            Tsung-Dao Lee Institute, Shanghai 200240, People's Republic of China\\
        \and
            School of Physics and Astronomy, Sun Yat-sen University, 2 Daxue Road, Tangjia, Zhuhai, 519082, People's Republic of China
\vs\no
   {\small Received~~20xx month day; accepted~~20xx~~month day}}

\abstract{ The pairwise velocity  generating function $G$ has deep connection
with both the pairwise velocity probability distribution function
and modeling of redshift space distortion (RSD). Its implementation  into
RSD modeling is often faciliated by  expansion into series of pairwise
velocity moments $\langle v_{12}^n\rangle$. Motivated by the logrithmic transformation of the
cosmic density field, we investigate an alternative expansion into series
of pairwise velocity cumulants  $\langle v_{12}^n\rangle_c$ . We numerically evaluate the convergence
rate of the two expansions, with three $3072^3$ particle simulations of the CosmicGrowth N-body
simulation series. (1) We find that the cumulant expansion performs
significantly better, for all the halo samples and redshifts investigated. (2) For modeling RSD at $k_\parallel<0.1\hmpc$,
including only the $n=1,2$ cumulants is sufficient. (3) But for modeling RSD at
$k_\parallel=0.2\hmpc$, we need and only need the $n=1,2,3,4$
cumulants. These results provide specific requirements on RSD modeling
in terms of $m$-th order statistics of the large scale strucure. 
\keywords{cosmology: dark energy, dark matter, large-scale structure of universe}
}

   \authorrunning{Chen et al.}            
   \titlerunning{Generating function and its implications on redshift space distortion modeling}  

   \maketitle
%
%
\section{Introduction}           
\label{sect:intro}

\label{sect:intro}

One of the most important issues in cosmology is to interpret the cosmic acceleration \citep{Riess1998,Perlmutter1999}.
Both dark energy and modified gravitational theories can produce the same expansion history.
Yet, they predict the different growth histories of the structure. 
Therefore, in observation, one can distinguish them by testing the structure growth rate $f(z)\sigma_8(z)$ through redshift-space distortion (RSD) \citep{Peebles1980,Kaiser1987,Scoccimarro2004}.
The observed position of galaxy in redshift space will be distorted by its peculiar velocity along the line of sight due to the Doppler shift. This RSD effect turns the isotropic distributed pattern of galaxies in real space into the anisotropic one in redshift space. 
Since peculiar velocity directly reflects the structure growth, by modeling the mapping from real space to redshift space, the peculiar velocity information can be extracted and used to constrain the cosmology.

Over the past decades, RSD has been proved to be a very powerful cosmological probe and adopted in many observational projects, such as 2dFGS \citep{Peacock2001,Hawkins2003}, SDSS \citep{Tegmark2006,Reid2012,Samushia2012,Tojeiro2012,Chuang2013}, 
VVDS\citep{Guzzo2008}, WiggleZ \citep{Blake2011}, 6dFGS \citep{Beutler2012,Johnson2014}, GAMA \citep{Simpson2016}, VIPERS \citep{delaTorre2013,Pezzotta2017,Mohammad2018}, FastSound \citep{Okumura2016},
BOSS \citep{White2015,Howlett2015,Li2016,Alam2017} and eBOSS \citep{Tamone2020,Bautista2020}. 
In near future, the ongoing and upcoming dark energy surveys like DESI, PFS,  Euclid, SKA, WFIRST (e.g. \citet{DESICollaboration2016,Amendola2018,Abdalla2015,Spergel2015}) will have the ability to constrain the structure growth rate at $\sim 1\%$ or even higher accuracy level. 
However, this target precision presents a sever challenge to the RSD modeling.

The difficulties of accurate RSD modeling come from three key ingredients.
(1) One is the mapping between real space and redshift space \citep{Peebles1980,Scoccimarro2004}. 
The mapping is nonlinear. For example, the redshift space 2-pt correlation function is determined by
not only the two-point correlation function in real space, but all the
$n$-th order correlation functions. It is also nonlocal, that real space
clustering at other scales can have significant contribution to a given
scale in redshift space.
(2) One is the nonlinear evolution of the
matter/halo density and velocity field, a long standing challenge in
modern cosmology. 
(3) The third is the nonlinear (and nonlocal) galaxy-halo-matter 
relation in not only the position space, but the whole phase space \citep{Desjacques2018,Huterer2018,Chen2018,Zhang2018}.
RSD models usually treat the redshift space correlation function or power spectrum 
as a expansion to a series of the density and velocity field statistics in real space.
For example, the distribution function approach \citep{Seljak2011} expresses the redshift space density in terms of series of summation of velocity moments, then obtain  the redshift space power spectrum from the correlators between the Fourier components of these moments.
\cite{Okumura2012a,Okumura2012b} investigate the contribution of each correlator in N-body simulations and give a conclusion that the accurate measurement of the redshift space power spectrum to $k\simeq 0.2 \hmpc$ at $z=0$ and $k\simeq 0.3\hmpc$ at $z=2$ require 6th order moment statistics to be taken into account.
The Fourier streaming model \citep{Vlah2019} expand the redshift space power spectrum with cumulant theorem. 
\cite{Chen2020} compare the moment expansion approach and the Fourier streaming model in N-body simulation halo samples. 
They conclude that the expansions have good agreement with the power spectrum at the percent level when third order velocity statistics is taken into account except those close to line of sight direction, while the forth order will break this agreement for $k>0.2\hmpc$.
Generally, existing models treat the large-scale velocity with perturbation theory then add the small scale Finger-of-God effect induced by the random motion in small scale, or assume a certain type of velocity distribution.
These approaches will mix all the non-linear effects together and make it difficult to quantify the influence of each individually.

In this paper, we take a step back from these works
and restrict our study to the first ingredient. The question that we
aim to ask is that, to accurately describe the real space-redshift
space mapping, what LSS statistics must be included.
As known in the
literature \citep{Scoccimarro2004}, the mapping is fully determined by the pairwise velocity
generating function $G$, this question then reduces to (1) what
expansion shall we adopt to describe $G$, and (2)  which order of
pairwise velocity moments shall we include in the expansion. 

In our previous work \citep{Zhao2021}, we directly evaluated the generating function $G$ at redshift $z=0$ in dark matter field. We also proposed a new RSD statistics $P^s(k_\parallel,r_\bot)$ which is more convenient to evaluate in the context of $G$.
In this work, we present the more comprehensive investigations to generating function,
including the halo mass and redshift dependence. 
We push the redshift to $z=1.5$ which is close to the interest of DESI, PFS, Euclid and SKA. 
And most importantly, we quantify the contribution from individual moments to $G$ and evaluate its impact to the hybrid statistics $P^s(k_\parallel,r_\bot)$.
Furthermore, we also investigate the influence of Gaussian and exponential as the pairwise velocity PDF to the measurement of generating function $G$.
\cite{Zhang2013} provides a method to decompose the peculiar velocity in different components with different features, which can help us to better understanding the peculiar velocity field and RSD modeling.
We also use the similar method in this work, to investigate the contributions and behaviors of $G$ for the different components.

We organize this paper as follows.
In \S 2, we provide a brief review of RSD modeling and its relation with the pairwise velocity moment generating function. 
Then we derive two independent approaches to measure the moment generating function in simulation.
\S 3 introduces the simulation and halo catalogs we adopt for numerically evaluation of the related quantities.
The main results are presented in \S 4.
Finally, \S 5 summarizes our major findings.

\section{Pairwise velocity generating function and RSD modeling} \label{sec:theory}

Comoving peculiar velocity ${\bf v}$ of a galaxy adds a Doppler redshift on
top of the cosmological redshift, $z^{\rm
  obs}=z+v_\parallel/c$. Here $v_{\parallel}={\bf v}\cdot \hat{x}$ is
the velocity component along the line of sight $\hat{x}$. Therefore
the observed position ${\bf s}$ of the 
galaxy in the redshift space is changed with respect to
its real space position ${\bf x}$,
\ba
    \mathbf{s}=\mathbf{x}+\frac{\mathbf{v} \cdot \hat{x}}{H(z)} \hat{x}=\mathbf{x}+\frac{v_{\|}}{H(z)} \hat{x}\ .
\ea
Here  $H(z)$ is the Hubble parameter at redshift $z$. For
brevity we will neglect $H$ in the denominator, so ${\bf v}$ hereafter
should be interpreted as ${\bf v}/H$. The
redshift space galaxy number density is then,
\ba
n^s({\bf s}) =\bar{n}(1+\delta^s({\bf
s}))=\sum_\alpha \delta_{3D}\left({\bf s}-\left[{\bf
    x}_\alpha+v_{\parallel,\alpha}\hat{x}_\alpha\right]\right)\ .
\ea
The sum is over all galaxies ($\alpha=1,2\cdots$) considered. The
Fourier transform of the overdensity $\delta^s$ is then
\ba
\bar{n}\left[\delta^s({\bf k})+(2\pi)^3\delta_{3D}({\bf k})\right]=\sum_\alpha \exp\left(i{\bf k}\cdot \left[{\bf
    x}_\alpha+v_{\parallel,\alpha}\hat{x}_\alpha\right]\right)\ .
\ea

\subsection{Power spectrum based models}
The redshift space power spectrum  $P^s({\bf k})$ is defined through
\ba
\langle \delta^s({\bf k})\delta^s({\bf k}^{'})\rangle=(2\pi)^3\delta_{3D}({\bf
  k}+{\bf k}^{'})P^s({\bf k})\ .
\ea
We then obtain 
\ba
\label{eqn:discretePS}
\bar{n}^2V\left(P^s({\bf k})+(2\pi)^3\delta_{3D}({\bf k})\right)=\left\langle \sum_{\alpha\beta}e^{ik_{\parallel}v_{\alpha\beta}} e^{i{\bf
    k}\cdot{\bf r}^{'}_{\alpha\beta}}\right\rangle\ .
\ea
Here we have adopted a fixed line of sight. $v_{\alpha\beta}\equiv
v_{\parallel,\alpha}-v_{\parallel,\beta}$. ${\bf r}^{'}\equiv {\bf
  x}_\alpha-{\bf x}_\beta$.  In the continuum limit, the above result
reduces to the more familiar form, 
\ba
\label{eqn:PS}
    P^{s}(\mathbf{k})=\int \left(\left\langle(1+\delta_{1})\left(1+\delta_{2}\right) e^{i k_{\|} v_{12}}\right\rangle_{\mathbf{r}'}-1\right) e^{i \mathbf{k} \cdot \mathbf{r}'} d^{3} \mathbf{r}'\ .
\ea 
in which $\delta_i\equiv \delta(\mathbf{x}_i)(i=1,2)$, $\mathbf{r}'\equiv\mathbf{x_1}-\mathbf{x}_2$, $v_{12}\equiv v_{\|}(\mathbf{x}_1)-v_{\|}(\mathbf{x}_2)$.
$\langle\cdots\rangle$ denotes the ensemble average. 
The subscript means the ensemble average is taken at a fixed pair
separation $\mathbf{r}'$. 

The above results are widely known in the literature
(e.g. \citet{Scoccimarro2004}).  Several models of RSD are based upon Eq. \ref{eqn:PS}, or
Eq. \ref{eqn:discretePS} or its
equivalent forms
(e.g. \citet{Scoccimarro2004,Matsubara2008b,Taruya2010,Seljak2011,Okumura2012a,Zhang2013,Zheng2013,Zheng2016,Song2018,Zheng2019}).

\subsection{Correlation function based models}
The redshift space correlation function is also modelled with the
streaming model \citep{Peebles1980}, 
\ba
    1+\xi^{s}(\mathbf{r}=(r_{\|}, \mathbf{r}_{{\bot}})) 
    =\int\left(1+\xi (\mathbf{r}'=(r_{\|}', \mathbf{r}_{{\bot}}))\right) p(v_{12} \mid \mathbf{r}'=(r_{\|}', \mathbf{r}_{{\bot}})) d r_{\|}'\ ,
    \label{eqn:streaming}
\ea
where $\mathbf{r}_\bot$ is the component of the separation in the
perpendicular direction to the line of sight. $p(v_{12}|{\bf r})$ is
the pairwise velocity PDF at separation ${\bf r}$. 

Eq.\ref{eqn:streaming} is exact. Nevertheless, $p(v_{12})$ is poorly
understood in theory and approximations of it are inevitable in
practice. The Gaussian steaming model \citep{Reid2011} 
takes the assumption $p(v_{12})$ distributes as the Gaussian function
with a non zero mean $\langle v_{12}\rangle$ and dispersion
$\sigma_{12}$. A further problem is that,  it is difficult to find a
suitable parametric form for $p(v_{12})$\citep{Fisher1995,Sheth1996,Juszkiewicz1998,Scoccimarro2004,Tinker2007,Bianchi2015,Bianchi2016,Kuruvilla2018,Cuesta-Lazaro2020}.

\subsection{Pairwise velocity generating function and RSD modeling}
The above two statistics can be unified by the pairwise velocity 
generating function\citep{Scoccimarro2004}, 
\ba
    G(k_{\|},\mathbf{r})=\frac{\langle(1+\delta_1)(1+\delta_2)e^{ik_{\|}v_{12}}\rangle}{1+\xi(r)},
\ea
where $\xi(r)=\langle\delta_1\delta_2\rangle$ is the two point correlation function in real space. 
One can verify that $G$ is  the generating function of the pairwise velocity,
\ba
    \langle v_{12}^{m}\rangle \equiv
    \frac{\langle(1+\delta_{1})(1+\delta_{2})
      v_{12}^{m}\rangle}{1+\xi(r)}=\left.\frac{\partial^{m}
        G}{\partial(i k_{\|})^{m}}\right|_{k_{\|}=0}, m \geq 1\ .
\ea
For the discrete distribution, the generating function should be defined by
and evaluated through 
\ba
G(k_{\|},\mathbf{r})\equiv \frac{\langle \sum_{\alpha\beta}\exp(ik_{\parallel}v_{\alpha\beta})\rangle_{{\bf r}_{\alpha\beta}={\bf r}}}{\langle
  \sum_{\alpha\beta}\rangle_{{\bf r}_{\alpha\beta}={\bf r}}}\ .
\ea
Here the ensemble average is over pairs with separation ${\bf
  r}_{\alpha\beta}={\bf r}$.  When $r\rightarrow \infty$ that we can
neglect spatial correlations in the density and velocity fields, 
\ba
G(k_\parallel,r\rightarrow \infty)\equiv  G_\infty=\left\langle
e^{ik_\parallel v_\alpha}\right\rangle^2\ .
\ea
This quantity is positive, and describes the Finger of God effect (e.g. \citet{Zhang2013,Zheng2013}). 

The pairwise velocity generating function plays an important role in RSD modelling.
\begin{itemize}
\item First, it determines the redshift power spectrum in Fourier space,
\ba
    P^{s}(\mathbf{k})=\int \left[\left(1+\xi(r^{'})\right) G(k_{\|}, \mathbf{r}^{'})-1\right] e^{i \mathbf{k} \cdot \mathbf{r}^{'}} d^{3} \mathbf{r}^{'}\ .
    \label{eq:psg}
\ea
\item Second, it determines the pairwise velocity PDF and therefore the RSD modelling in
  configuration space.
\ba
    p(v_{12} \mid \mathbf{r})=\int G(k_{\|}, \mathbf{r}) e^{i k_{\|} v_{12}} \frac{d k_{\|}}{2 \pi}\ .
    \label{eqn:pwpdf}
\ea
\item The above relations are well known in the literature
(e.g. \citet{Scoccimarro2004,Taruya2010,Desjacques2018b}), but have not fully incorporated in RSD
modelling. Furthermore, we can define a hybrid statistics
$P^s(k_{\|},r_{\bot})$. 
By multiplying both sides of Eq.\ref{eq:psg} by $\int\exp(-i\mathbf{k}_{\bot}\cdot\mathbf{r}_{\bot})d^2\mathbf{k}_{\bot}/(2\pi)^2$, we obtain
\ba
\label{eqn:hybrid}
    P^s(k_{\|},r_{\bot})=\int
    \left[\left(1+\xi(r)\right)G(k_{\|},\mathbf{r})-1\right]e^{ik_{\|} r_{\|}}d
    r_{\|}\ .
\ea
This is neither the correlation function nor the power spectrum. But
this hybrid statistics has some attractive features. (1) Since
$G(k_\parallel=0)=1$, $P^s(k_{\|} =
0, r_{\bot}) =\int_{-\infty}^{\infty}\xi(r_\parallel,r_\perp)
dr_{\parallel}=w_p(r_{\bot})$. Namely, the $k_{\|} = 0$ mode equals the projected correlation function $w_p$, 
\footnote{The projected correlation function $w_p(r_\perp)$ is often redefined as $w_p(r_\perp)/r_\perp$
to make it dimensionless.} therefore, it is unaffected by RSD, which is only constrained to $k_{\|}\neq 0$ modes. 
This is an advantage that $P^s({\bf k})$ also share. 
But $\xi^s$ does not have this advantage,  since $\xi^s(r_\parallel,r_\perp)$ of
all configurations are affected by RSD. (2)  Within the context of RSD modelling
with the generating function $G$, this is the most straightforward to
numerically implement, since only one integral over $r_\parallel$ is
needed. (3) In the measurement, it is also straightforward to convert
from the measurement of correlation function, which has better
handling over survey masks and varying line of sight.
\end{itemize}

\subsection{Moment and cumulant expansion of  the generating function}
One intrinsic advantage is that $G$ can be naturally
Taylor expanded with physically meaningful Taylor coefficients. This
can be implemented either  with the moment expansion or with the cumulant
expansion. 

\subsubsection{Moment expansion}
The moment expansion directly expands $G$ into its Taylor expansion
series, 
\ba
\label{eqn:taylor}
    G(k_{\|}, \mathbf{r})&=&1-\sum_{m\geq 1}(-1)^{m-1} \frac{\langle
    v_{12}^{2m}\rangle}{(2m)!}k_{\|}^{2m} 
    +i\sum_{m\geq 1}(-1)^{m-1} \frac{\langle
      v_{12}^{2m-1}\rangle}{(2m-1)!}k_{\|}^{2m-1}\nonumber \\
&=&1+ i\langle v_{12}\rangle
    k_{\|}-\frac{1}{2}\langle v_{12}^{2}\rangle k_{\|}^{2}
    -\frac{1}{6}i\langle v_{12}^{3}\rangle
    k_{\|}^{3}+\frac{1}{24}\langle v_{12}^{4}\rangle k_{\|}^{4}+\cdots \ .
\ea
The convergence rate of Eq.\ref{eqn:taylor} is decided by the
coefficients of pairwise velocity moments. Through numerical
simulations, we can robustly quantify the impact of individual terms
and determine the moments which must be included to reach the desired
accuracy in RSD.

\subsubsection{Cumulant expansion}
Eq. \ref{eqn:taylor} is not the only way of expanding $G$ in velocity
moments. Instead we can Taylor expand $\ln G$ in power series of
$k_\parallel$. The expansion coefficients turn out to be the pairwise
velocity cumulants $\langle v_{12}^m\rangle_c$.  \cite{Scoccimarro2004}
already pointed out $\ln G$ as the cumulant generating function, but
did not specify the cumulant expansion coefficient as $\langle
v_{12}^m\rangle_c$. Therefore we provide a proof here. Furthermore, we
find that such relation is
connected to the widely adopted logarithmic transformation of the
cosmic density field.  

Defining an auxiliary field
\ba
    y\equiv \ln(1+\delta)-\langle\ln(1+\delta)\rangle\ ,
\ea
and setting $\lambda=ik_{\|}$, we have
\begin{eqnarray}
    G(\lambda \mid \mathbf{r})  &\equiv& \frac{\langle(1+\delta_{1})(1+\delta_{2}) \exp (\lambda v_{12})\rangle}{\langle(1+\delta_{1})(1+\delta_{2})\rangle}\nonumber \\ 
    &=&\frac{\langle\exp \left[(y_{1}+y_{2})+\lambda v_{12}\right]\rangle}{\langle(1+\delta_{1})(1+\delta_{2})\rangle} \nonumber\\ 
    &=&\frac{1}{1+\xi(r)}\exp \left[\sum_{n \geq 2} \frac{\langle((y_{1}+y_{2})+\lambda v_{12})^{n}\rangle_{c}}{n !}\right] \nonumber\\ 
    &=&\exp \left[\sum_{m \geq 1} \frac{c_{m}(\mathbf{r})}{m!}
        \lambda^{m}\right]\ .
\end{eqnarray}
Here,
\ba
    c_{m} \equiv m! \sum_{n \geq 2, n \geq m} \frac{C_{n}^{n-m}}{n !}\frac{\left\langle\left(y_{1}+y_{2}\right)^{n-m} v_{12}^{m}\right\rangle_{c}}{1+\xi(r)}\ .
\ea
Furthermore,  we find 
\begin{eqnarray}
    c_{1} &=&\langle v_{12}\rangle \equiv \langle v_{12}\rangle_c\ , \no \\ 
     c_{2} &=&\langle (v_{12}-c_1)^2\rangle
                   \equiv \langle v^2_{12}\rangle_c\no \\ 
    c_{3}  &=&\langle (v_{12}-c_1)^3\rangle\equiv \langle v^3_{12}\rangle_c\no\\ 
    c_{4}&=&\langle (v_{12}-c_1)^4\rangle-3 \langle (v_{12}-c_1)^2\rangle\equiv \langle
                     v^4_{12}\rangle_c\ ,\cdots
\end{eqnarray}
Namely the cumulant expansion coefficient $c_m$ is the pairwise
velocity cumulant $\langle v_{12}^m\rangle_c$.  
Then we obtain the cumulant expansion of pairwise velocity generating
function, 
\ba
  \ln G(k_{\|}, \mathbf{r}) &=& -\sum_{m \geq 1}
  (-1)^{m-1}\frac{\langle v_{12}^{2m}\rangle_c}{(2m)!} k_{\|}^{2 m} 
  + i \sum_{m \geq 1} (-1)^{m-1} \frac{\langle v_{12}^{2 m-1}\rangle_c}{(2m-1)!}
  k_{\|}^{2 m-1}\ .
  \label{eqn:cumulant}
\ea
Namely $\ln G$ is the cumulant generating function of pairwise
velocity, versus $G$ as the moment generating function of pairwise
velocity. We may have expected this correspondence from the
moment/cumulant  generating function of the density field. 

We may expect that the above expansion converges faster than the
expansion with Eq. \ref{eqn:taylor}. The reason is that the density
field is close to lognormal and the velocity field is close to
Gaussian. Under such condition, only the $k_\parallel$ and
$k_\parallel^2$ terms exist in Eq. \ref{eqn:cumulant}.  This results
in a Gaussian pairwise velocity PDF, and corresponds to the Gaussian
streaming model of correlation function. Numerical evaluation later
indeed shows that the expansion of Eq. \ref{eqn:cumulant} indeed
converges faster than that of Eq. \ref{eqn:taylor}. Nevertheless, we
find $k_\parallel^{3,4}$ terms are non-negligible at $k\ga 0.2\hmpc$,
implying further improvement over the Gaussian streaming
approximation. 

\subsection{Peculiar velocity decomposition}
As mentioned above, moment generating function determines the pairwise velocity PDF, and vice versa. There are lots of models based on the pairwise velocity PDF assuming some specific forms of pairwise velocity PDF, such as Gaussian distribution \citep{Reid2011}, exponential distribution \citep{Sheth1996} and so on.
Here we investigate the influence of both Gaussian and exponential approximations towards generating function.
For brevity, here we only provide one point statistics of velocity PDF, 
instead of the more complicated two point statistics of pairwise velocity PDF.
The statistics from simulation prefer a mixture of Gaussian and exponential pairwise velocity PDF.
At sufficiently large scale with low speed, it is close to Gaussian distribution, 
\ba
    p_\mathcal{G}(v)=\frac{1}{\sqrt{2\pi \sigma_\mathcal{G}^2}}\exp({-v^2/2\sigma_\mathcal{G}^2})\ ,
\ea
yet at small scale with severe random motions, it turns to exponential distribution,
\ba
    p_\mathcal{E}(v)=\frac{1}{\sqrt{2 \sigma_\mathcal{E}^2}}\exp({-\sqrt{2}|v|/\sigma_\mathcal{E}})\ .
\ea
Here $\sigma_{\mathcal{G,E}}$ is the pairwise velocity dispersion for Gaussian/Exponential components, and $\sigma_\mathcal{G}^2+\sigma_\mathcal{E}^2=\sigma^2$.
The corresponding Fourier transformations are,
\ba
    \mathcal{G}=\exp({-\sigma_\mathcal{G}^2 k^2/2})\ , \mathcal{E}=\frac{1}{\sigma_\mathcal{E}^2 k^2/2+1}\ .
\ea
Assuming the Gaussian part and exponential part are independent with each other, 
the generating function could be written as,
\ba
    G\approx \mathcal{G(\sigma_\mathcal{G})}\mathcal{E(\sigma_E)}
    \label{eqn:gl_mgf}
\ea
Under these assumptions, the imaginary part of generating function, Im($G$), vanishes. 
Gaussian/exponential distributions determine the upper/lower limits of Re($G$).

Furthermore, \cite{Zhang2013} provide a method to decompose the peculiar velocity field into three parts, ${\bf v}_\delta$, ${\bf v}_B$, and ${\bf v}_S$.
${\bf v}_\delta$ is the over-density field correlated part. It dominates at the linear scale where $k\ll k_{NL}$ ($NL$ is short for ``non-linear'' scale), then vanishes due to the nonlinear evolution at small scale.
Differ from ${\bf v}_\delta$, the stochastic component ${\bf v}_S$ and rotational component ${\bf v}_B$ only reveal and dominant at the nonlinear scale.
\cite{Zheng2013} verified these theories in N-body simulation.
In this paper, we decompose the peculiar velocity into only density correlated (the deterministic) part, 
and the rest stochastic part (${\bf v}_S+{\bf v}_B$ in \cite{Zhang2013}). 
We denoted them with superscripts $L$ and $S$ respectively,
${\bf v}({\bf x})={\bf v}^L({\bf x})+{\bf v}^S({\bf x})$.
In Fourier space,
\ba
    \mathbf{v}^L(\mathbf{k})=-i\frac{H(z)\delta(\mathbf{k})W(k)}{k^2}\mathbf{k}\ .
    \label{eqn:vlin}
\ea
Here, the window function, 
\ba
    W(k)=\frac{P_{\delta\theta}(k)}{P_{\delta\delta}(k)}\ .
\ea
in which, $\theta=-\nabla\cdot{\bf v}$ is the divergence of the peculiar velocity.
Then the generating function can be expressed in,
\begin{eqnarray}
    G&=&\frac{\langle(1+\delta_1)(1+\delta_2)\exp{(ik_{\|}v)}\rangle}{1+\xi(r)}\nonumber \\
    &=&\frac{\langle(1+\delta_1)(1+\delta_2)\exp{(ik_{\|}v^L)}\exp{(ik_{\|}v^S)}\rangle}{1+\xi(r)}
\end{eqnarray}
If the density field is log-normal, assume $L$ and $S$ components are independent with each other, we have
\ba
    \ln G=\ln G^L + \ln G^S\ .
    \label{eqn:ls_mgf}
\ea
$G^L$ is expected to be approximately Gaussian, and the stochastic part $G^S$ should be close to exponential.
We can evaluate the convergence of both Eq.(\ref{eqn:gl_mgf}) and Eq.(\ref{eqn:ls_mgf}) in simulation.

\section{Simulation}\label{sec:data}

\begin{table}
    \centering
    \caption{Three sets of halo mass bins for J6610. 
  The mass unit is in $10^{12} M_\odot/h$. 
  $\langle M \rangle$ is the mean halo mass. 
  $N_h$ is the total halo number in corresponding halo mass bin.}
    \label{tab:halosets}
    \begin{tabular}{lcccc}
    \hline
    \hline
    Set ID & Mass Range & $\langle M \rangle$ & $N_h/10^4$ \\
    \hline
      $ A1(z=0.0) $ & $>10$ & 37.70 & 8.66  \\
      $    z=0.5\ $ & $>10$ & 30.03 & 6.70  \\
      $    z=1.0\ $ & $>10$ & 23.77 & 4.30  \\
      $    z=1.5\ $ & $>10$ & 20.39 & 2.53  \\
    \hline
      $ A2(z=0.0) $ & 1-10 & 2.67 & 69.23  \\
      $    z=0.5\ $ & 1-10 & 2.61 & 66.88  \\
      $    z=1.0\ $ & 1-10 & 2.51 & 59.86  \\
      $    z=1.5\ $ & 1-10 & 2.41 & 50.49  \\
    \hline
      $ A3(z=0.0) $ & 0.1-1 & 0.27 & 506.14  \\
      $    z=0.5\ $ & 0.1-1 & 0.27 & 523.57  \\
      $    z=1.0\ $ & 0.1-1 & 0.26 & 527.22  \\
      $    z=1.5\ $ & 0.1-1 & 0.26 & 508.97  \\
    \hline
    \end{tabular}
\end{table}

We numerically evaluate the generating function $G$ at various
$k_\parallel$ and $(r_\parallel,r_\perp)$, and the two expansion
series (Eq. \ref{eqn:taylor} \& \ref{eqn:cumulant}),  in a subset of the
CosmicGrowth simulations \citep{Jing2019}.  
The three simulations are run with a particle-particle-particle-mesh
($\mathrm{P^3M}$) code \citep{Jing2007}, boxsize $L_{\mathrm{box}}=600
\mpch$, and particle number $N_P=3072^3$. They adopt the identical
$\mathrm{\Lambda CDM}$ cosmology, with $\Omega_b=0.0445$,
$\Omega_c=0.2235$, $\Omega_{\mathrm{\Lambda}}=0.732$, $h=0.71$,
$n_s=0.968$ and $\sigma_8=0.83$.  
It has three realizations, denoted as J6610, J6611 and J6612 here.
The halo catalogs are first identified by a Friends-of-Friends (FoF)
algorithm, with the linking length $b=0.2$ times the mean
inter-particle separation. Then all unbound particles have been removed from the catalogs.
We select three different halo mass bins, labeled as A1, A2 and A3, at
four redshift snapshots, $z\simeq 0,0.5,1.0,1.5$. The mass range, mean
mass, and total number of each halo set for J6610 are listed in
Table~\ref{tab:halosets}. Specifications of  J6611 and J6612 are similar. 

We use the NGP method with $600^3$ grid points to construct the needed fields.
The grid size is $L_{\mathrm{grid}}=1 \mpch$.
For each grid, we measure $\alpha_i=\sum_\gamma \cos(k_{\|}v_{\|,\gamma})$, 
$\beta_i=\sum_\gamma \sin(k_{\|}v_{\|,\gamma})$ and $p_i^n=\sum_\gamma
v_{\|,\gamma}^n$, $n=0,1,2,3,\dots$. Notice that 
$p^0_i=\sum_\gamma=(1+\delta_i)$. The summation $\sum_\gamma$ is over
all particles nearest to the $i$th grid point. The real and imaginary
part of the generating function are evaluated separately by the
following relation
\ba
\label{eqn:GS}
    G(k_\parallel,{\bf r})
    =\frac{\langle\beta_1\beta_2+\alpha_1\alpha_2\rangle_{\bf r}}{\langle
      p_1^0 p_2^0\rangle_{\bf r}}
    +i\frac{\langle\alpha_1\beta_2-\beta_1\alpha_2\rangle_{\bf r}}{\langle p_1^0 p_2^0\rangle_{\bf r}} \ .
\ea
The pairwise velocity moments is given by
\ba
\label{eqn:v12S}
\langle v_{12}^m\rangle=\frac{\langle \sum_{n=0}^m
      C_m^{m-n}(-1)^n p_1^n p_2^{m-n} \rangle_{\bf r}}{\langle p_1^0 p_2^0\rangle_{\bf r}}\ .
\ea
One  thing to notice is that,  the r.h.s. of Eq. \ref{eqn:GS} \&
\ref{eqn:v12S} means that we can utilize FFT to speed up the
computation. For each $k_\parallel$, $8$ FFTs are needed to evaluate
$G$ of all ${\bf r}$ pairs, and $\geq 3$ FFTs for $\langle
v_{12}^m\rangle$. Nevertheless, since we only investigate a dozen
${\bf r}$ specifications,  we instead measure the above quantities by
directly counting the pairs with 
fixed separation $r_{\|}$ and $r_\bot$ values. We can choose the
Cartesian $x$, $y$, $z$ axes of the simulation box as the line of
sight, so for each simulation we have 3 independent measurements. With
3 independent simulation realizations, we have $9$ independent
measurements and we can then estimate the errorbars of the measured
properties. 

In order to obtain the deterministic and stochastic components of halo peculiar velocity field, first we measure $\mathbf{v}^L(\mathbf{k})$ from Eq. \ref{eqn:vlin}. When obtaining the quantities in Eq.\ref{eqn:vlin}, $512^3$ number of grid points are adopted to construct the necessary fields. 
Then we do inverse FFT to obtain $\mathbf{v}^L$ in configuration space.
\cite{Chen2018} has verified that the large scale velocity bias between halo and dark matter is unity in N-body simulation,
and \cite{Zhang2018} provides the possible explanation. 
Thus here we can treat the deterministic velocity component of halos as the underlying dark matter's 
, $\mathbf{v}_h^L=\mathbf{v}^L$.
Finally, we obtain the stochastic component by $\mathbf{v}_h^S=\mathbf{v}_h-\mathbf{v}^L$.

\section{Numerical Results and implications}
\label{sec:result}

\begin{figure}
    \centering
    \includegraphics[width=12cm, angle=0]{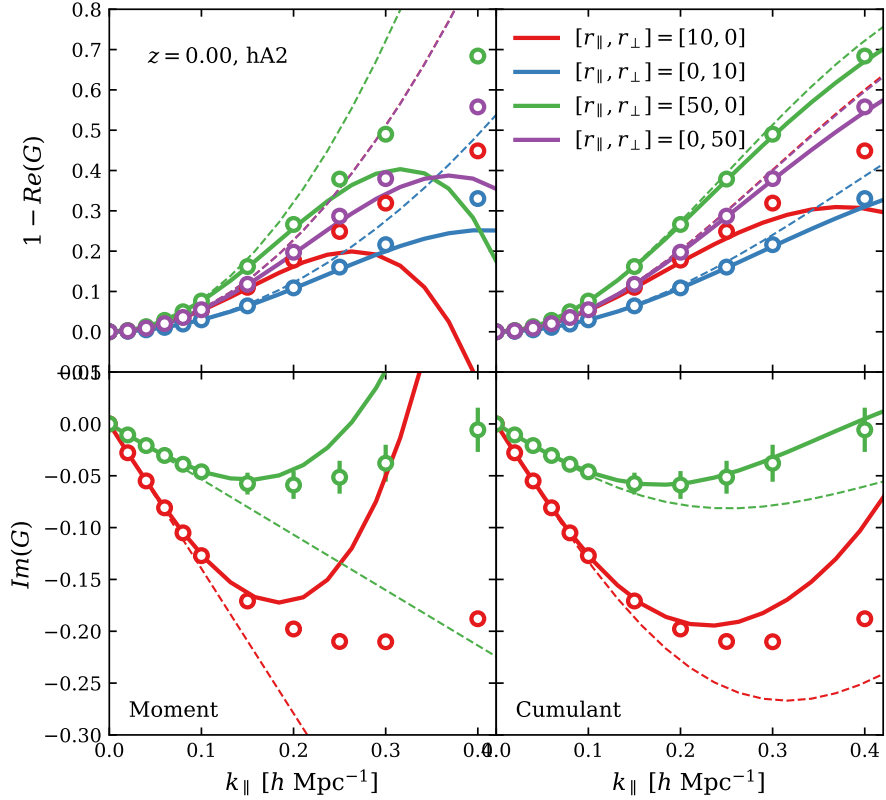}
    \caption{
    The pairwise velocity generating function $G$ at $z=0$, 
    for the halo set A2 $(10^{12}M_\odot/h<M<10^{13}M_\odot/h)$. 
    Data points (with errorbars) are directly measured from the $3$ simulation realizations 
    and the errorbars are r.m.s of the $9$ independent measurements 
    ($3$ simulation realization $\times$ three directions). 
    Top/bottom panels correspond to the real/imaginary part of $G$ respectively. 
    Left/right panels correspond to the results of moment/cumulant expansions. 
    The dash lines cut off at the leading order terms ($\langle v_{12}^{1,2}\rangle$,$\langle v_{12}^{1,2}\rangle_c$),
    while the solid lines include the next-to-leading order terms 
    ($\langle v_{12}^{1,2,3,4}\rangle$,$\langle v_{12}^{1,2,3,4}\rangle_c$). 
    The major finding is that the cumulant expansion works significantly better than the moment expansion. 
    The leading order approximation is excellent at $k\leq 0.1\hmpc$. 
    Including $\langle v_{12}^{3,4}\rangle$, the cumulant expansion is excellent at
    $k<0.2\hmpc$ for all $(r_{\|},r_\bot)$ configurations. 
    Furthermore, for $r_{\|}\geq 20 \mpch$, 
    it is excellent to $k\sim 0.4\hmpc$.
    Bottom panels (Im$G$) does not show the configurations with $r_{\|}=0$, 
    for which Im$G$=0 due to the $v_{\|} \leftrightarrow \- v_{\|}$ symmetry.
    }
    \label{fig:cvg}
\end{figure}

\begin{figure}
    \centering
    \includegraphics[width=12cm, angle=0]{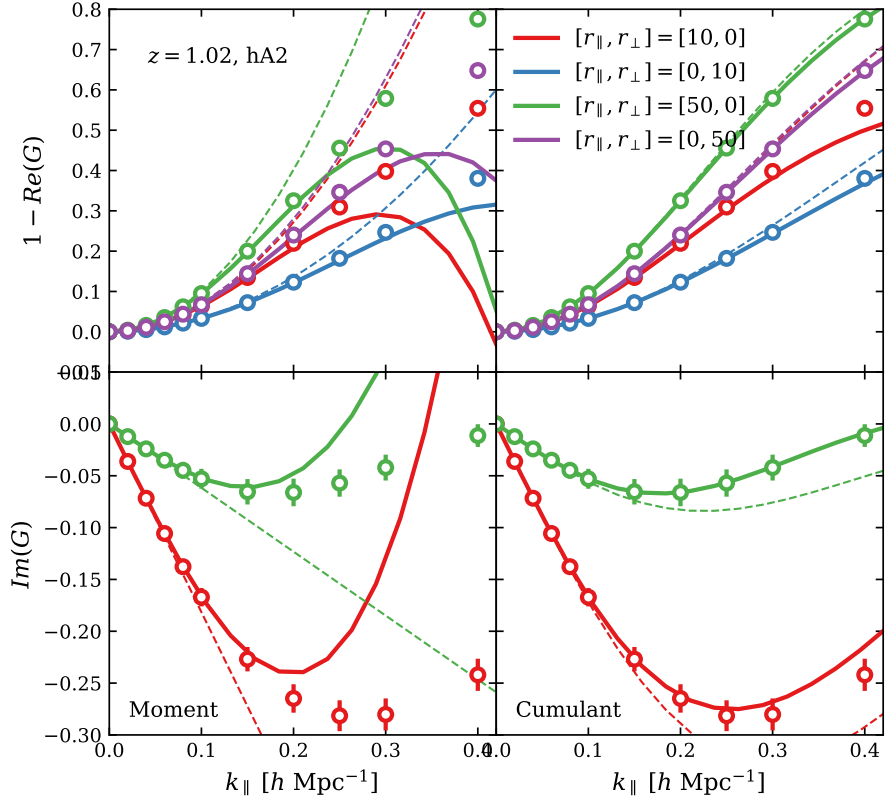}
    \caption{Similar to Fig. \ref{fig:cvg} but for redshift $z\approx 1$.}
    \label{fig:cvg2}
\end{figure}

The generating function $G\equiv G(k_{\|},r_{\|},r_\bot)$
depends on $k_{\|}$, $r_{\|}$, $r_\bot$ as well as redshift and halo
mass. We are not able to show the results of all possible
combinations. Instead, we will mainly show the result of mass bin A2.
To the same order of moment/cumulant expansion, the accuracy is
slightly better for A1, which is less affected by small scale
nonlinearities due to larger smoothing associated with the halo
mass/size. But since A1 has at least a factor of $10$ smaller halo numbers,
the measurements are more noiser.  In contrast, the accuracy for A3 is
slightly worse than A2, while the measurement noise is
smaller. Therefore in the main text we only show A2 as the
intermediate case. For the redshifts, we mainly show the case of $z=0$
and when necessary, the case of $z=1$. For the wavenumber $k$, the
primary target is $k=0.2\hmpc$, matching the capability of stage IV
projects. But since stage V projects have the capability to reach
$k\sim 0.5\hmpc$, we will also show the results of $k>0.2\hmpc$ in the
main text. 

\subsection{$\langle v_{12}^{3,4}\rangle$ terms must be included}
Fig. \ref{fig:cvg} shows $G$ as a function of $k_\parallel$,  at
$z=0$ and for
$(r_{\|},r_\bot)=(10,0),(0,10), (50,0), \& (0,50)$ (unit in
$\mpch$). We compare the leading order expansion to the simulated $G$.  As a reminder,
the leading order moment expansion is $G\simeq 1+i\langle
v_{12}\rangle k_\parallel-\langle
v_{12}^2\rangle k^2_\parallel/2$. The leading order cumulant expansion
is $G\simeq \exp(1+i\langle
v_{12}\rangle k_\parallel-\langle
v_{12}^2\rangle_c k^2_\parallel/2)$. All the coefficients
($\langle v_{12}^n\rangle$ and $\langle v_{12}^n\rangle_{c}$) are measured from the same simulation. The moment expansion becomes
inaccurate at $k_\parallel=0.1\hmpc$, especially for the imaginary part of
$G$. The cumulant expansion remains accurate at
$k_\parallel=0.1\hmpc$. Since the cumulant expansion up to leading
order is equivalent to a Gaussian $p(v_{12}|r_\parallel,r_\perp)$, this
explains the validity of Gaussian streaming model \citep{Reid2011} at sufficiently
large scale. However, at $k_\parallel\sim 0.2\hmpc$, the leading order
approximation results into significant error in the imaginary part of
$G$. 

\begin{figure}
    \centering
    \includegraphics[width=12cm, angle=0]{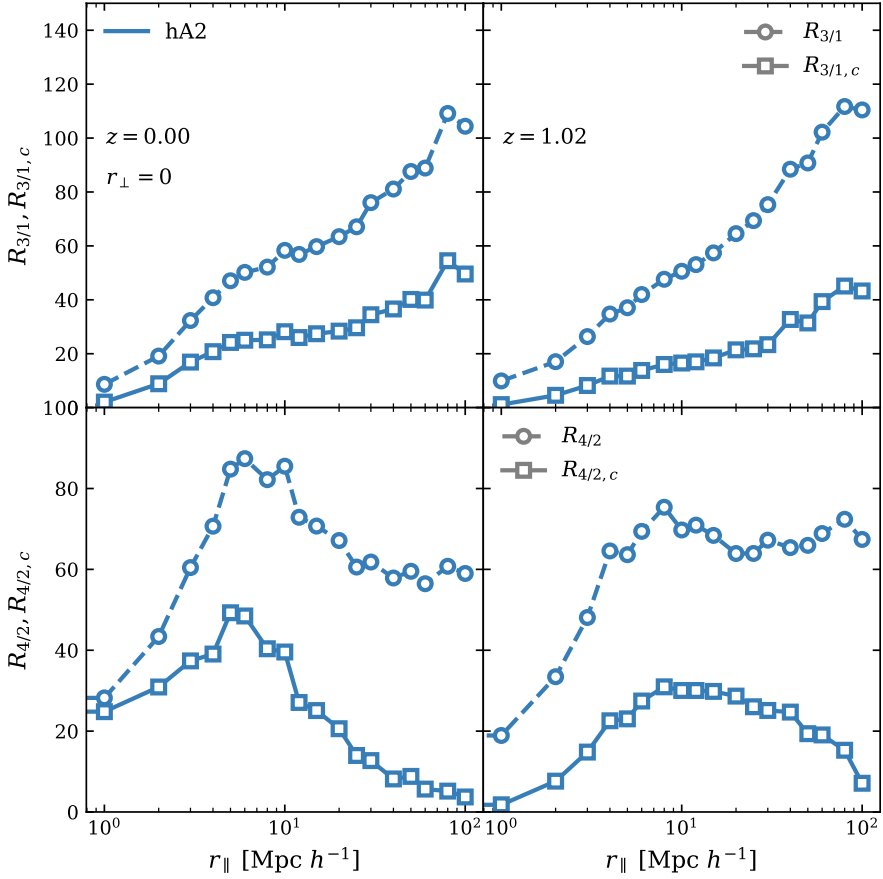}
    \caption{$R_{3/1}\equiv \langle v_{12}^3\rangle/\langle
      v_{12}\rangle$ and $R_{4/2}\equiv \langle v_{12}^4\rangle/\langle
      v_{12}^2\rangle$. These two determine the relative importance of
    the next-to-leading order terms in the moment expansion. For the
    cumulant expansion, the corresponding quantities are $R_{3/1,c}$
    and $R_{4/2,c}$ respectively. These results explain the necessity
    of including the next-to-leading terms in the generation function
    (and RSD). They also explain why the cumulant expansion is better
    than the moment expansion.}
    \label{fig:coef2}
\end{figure}

Therefore to improve the approximation accuracy at the target
$k_\parallel=0.2\hmpc$, we must include the next-to-leading order
terms in the expansion. Then the moment expansion becomes  $G\simeq 1+i[\langle
v_{12}\rangle k_\parallel-\langle
v_{12}^3\rangle k^3_\parallel/6]-[\langle
v_{12}^2\rangle k^2_\parallel/2-\langle
v_{12}^4\rangle k^4_\parallel/24]$. Nonetheless, the moment expansion
still fails at $k_\parallel\sim 0.2\hmpc$, especially for the
imaginary part. 

Including the next-to-leading order
terms, the cumulant expansion becomes $G\simeq \exp(i[\langle
v_{12}\rangle k_\parallel-\langle
v_{12}^3\rangle_c k^3_\parallel/6]-[\langle
v_{12}^2\rangle_c k^2_\parallel/2-\langle
v_{12}^4\rangle_c k^4_\parallel/24])$. This expansion is accurate at
$k_\parallel=0.2\hmpc$. It remains accurate even until
$k_\parallel\sim 0.4\hmpc$, unless $r_\perp\rightarrow 0$. 

The situation is similar at other redshifts (e.g. $z=1$,
Fig. \ref{fig:cvg2}). Therefore the first major result of this paper
is that, to accurately describe $G$ at $k_\parallel\sim 0.2\hmpc$, we have to include not only
$\langle v_{12}^{1,2}\rangle$, but also $\langle v_{12}^{3,4}\rangle$
into the model.  Since $G$ completely determines RSD, this also
implies that we must include $\langle v_{12}^{3,4}\rangle$
into the modeling of RSD. This will be challenging, since  $\langle
v_{12}^{3,4}\rangle$ themselves involve LSS correlations up to $6$-th
order ($\delta^2v^4$). 

We further check the origin of the above finding. 
The ratio of the $k_\parallel^3$ term to $k_\parallel$ term is
$R_{3/1}k_\parallel^2/6$ for the moment expansion, and
$R_{3/1,c}k_\parallel^2/6$ for the cumulant expansion. Here, $R_{3/1}\equiv \langle
v_{12}^3\rangle/\langle v_{12}\rangle$ and $R_{3/1,c}\equiv\langle
v_{12}^3\rangle_c/\langle v_{12}\rangle$. Fig. \ref{fig:coef2} shows
$R_{3/1}$ and $R_{3/1,c}$ for the case of $r_\perp=0$, which is among
the most difficult to model for the generating function and
RSD. $R_{3/1}$ and $R_{3/1,c}$ have typical values $\sim
10$-$100(\mpch)^2$. Therefore for $k_\parallel\ga 0.1\hmpc$, the
$k^3_\parallel$ term will become non-negligible comparing to the
$k_\parallel$ term. This problem does not alleviate toward large
separation, as we expect. In contrast, $R_{3/1}$ increases and the
problem becomes worse at large pair
separation. In fact,   at $r_\parallel\sim 100\mpch$, the moment
expansion to third order even fails to correctly predict the sign of Im$G$
for $k\geq 0.25\hmpc$.

The ratio
of the $k_\parallel^4$ term to $k_\parallel^2$ term is $R_{4/2}
k_\parallel^2/12$ for the moment expansion, and $R_{4/2,c}
k_\parallel^2/12$ for the cumulant expansion.  Here, $R_{4/2}\equiv \langle
v_{12}^4\rangle/\langle v_{12}^2\rangle$ and $R_{4/2,c}\equiv \langle
v_{12}^4\rangle_c/\langle v_{12}^2\rangle$.  The numerical results are
also shown in Fig. \ref{fig:coef2}. The worst inaccuracy of expanding to $4$-th
order occurs where $R_{4/2}$ ($R_{4/2,c}$) is largest. This happens at
$r_\parallel\sim 5\mpch$ and the typical value is $\sim 50(\mpch)^2$.
Notice that max$R_{4/2}< {\rm max}R_{3/1}$. Together with the extra
factor $1/2$ in the Taylor expansion, the relative correction is
significantly smaller in the real part of $G$ than that in the
imaginary part. 

\begin{figure}
    \centering
    \includegraphics[width=12cm, angle=0]{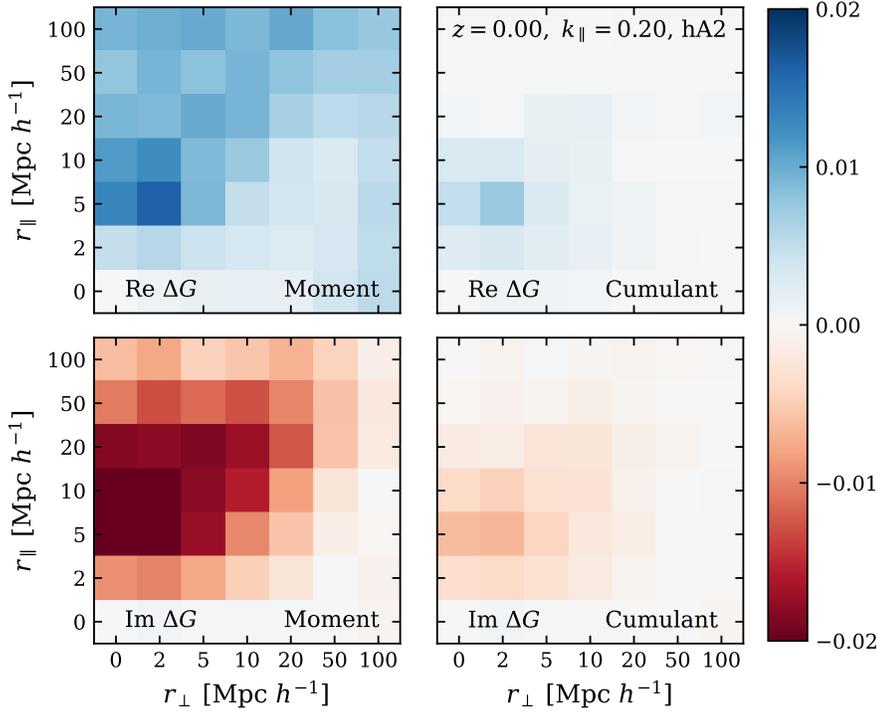}
    \caption{Accuracies of the moment expansion (left panels) and
      cumulant expansion (right panels) of the
      pairwise velocity  generating function $G$ at $k_\parallel=0.2\hmpc$. Both expansions keep the
      next-to-leading order terms, namely include all
      $k_\parallel^{1,2,3,4}$terms.  Top (bottom) panels are
      the results of real(imaginary) part of $G$. This comparison
      clearly shows that the cumulant expansion works significantly
      better than the moment expansion.  It achieves $|\Delta
      G|<0.01$ for all configurations of $r_\perp,r_\parallel$.  For brevity, we only show the
      comparison for the halo mass bin A2 at $z=0$. Results of other
      mass bins and redshifts are similar. }
    \label{fig:2ddiff}
\end{figure}

\subsection{Cumulant expansion is better}
Fig. \ref{fig:2ddiff} shows the errors by neglecting
$k_\parallel^{n>4}$ terms  in the moment/cumulant expansion, in the
$r_\perp$-$r_\parallel$ plane, for $k_\parallel=0.2\hmpc$. For the
whole range of interest ($r_\perp<100\mpch$, $r_\parallel<100\mpch$),
the cumulant expansion is better than the moment expansion.  The errors
are largest at $r_\perp\la 5\mpch$ and $r_\parallel\sim
5$-$10\mpch$. Nonetheless, $|\Delta G|\la 0.01$.

\begin{figure}
    \centering
    \includegraphics[width=12cm, angle=0]{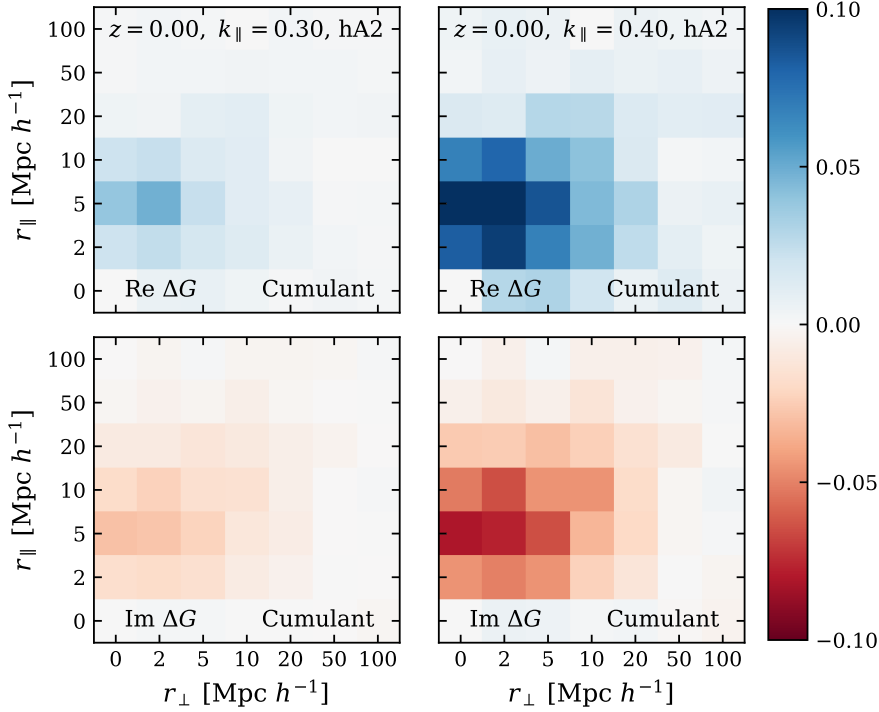}
    \caption{Similar to Fig. \ref{fig:2ddiff}, but only for the cumulant
  expansion at two different $k_z=0.3$ and $0.4$. \label{fig:0.30.4}}
\end{figure}

\begin{figure}
    \centering
    \includegraphics[width=12cm, angle=0]{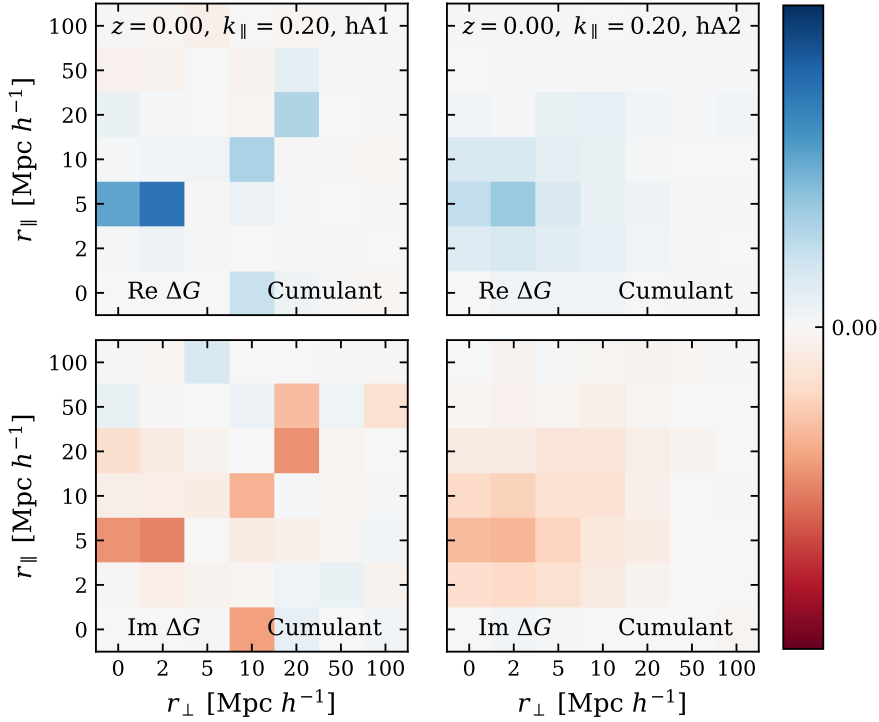}
    \caption{Similar to Fig. \ref{fig:0.30.4}, but for the two different halo catalogs A1 and A3 at fixed $k_z=0.2$. \label{fig:A1A3}}
\end{figure}

Fig. \ref{fig:0.30.4} shows the errors at $k_\parallel=0.3,0.4\hmpc$
for the cumulant expansion. The errors increase with $k_\parallel$, as
expected. Also as the case of $k_\parallel=0.2\hmpc$,  the largest error
occurs at $r_\perp\la 5\mpch$ and $r_\parallel\sim
5$-$10\mpch$ and max$|\Delta G|\sim 0.1$ for
$k_\parallel=0.4\hmpc$. Nonetheless, if we only use the region at
$r_\perp=20\mpch$,  the error in $G$ is reduced to $\sim 0.01$, even
for $k=0.4\hmpc$. Fig. \ref{fig:A1A3} shows the errors at
$k_\parallel=0.2\hmpc$, but for halo set A1 and A3. The cumulant
expansion is also excellent. 

Therefore the 
major results of this paper are
\ba
G&\simeq& \exp\left[-\frac{\langle v_{12}^2\rangle_c
  k_\parallel^2}{2}+i\langle v_{12}\rangle k_\parallel\right]\nonumber \\
&&\ {\rm for\ all\  } {\bf r}, {\rm but}\ \ k\leq 0.1\hmpc\ , \nonumber \\
&\simeq& \exp\left[-\frac{\langle v_{12}^2\rangle_c
  k_\parallel^2}{2}+\frac{\langle v_{12}^4\rangle_c
  k_\parallel^4}{24}+i\left(\langle v_{12}\rangle k_\parallel-\frac{\langle
v_{12}^3\rangle_c k_\parallel^3}{6}\right)\right] \nonumber \\
&&\ {\rm for\ all\  } {\bf r}, {\rm but}\ \ k\leq 0.2\hmpc\ ,\nonumber \\
&&\ {\rm or\ for\  r_{\perp}>20\hmpc} \ \&\   k\leq 0.4\hmpc\ .
\ea

\subsection{Induced errors in the RSD modelling}
Analysis above shows  that it's necessary to include at least 3rd and
4-th order pairwise velocity moments/cumulants in the modeling of generating
function at $k\ga 0.2\hmpc$. Inaccuracies in the generating function
modeling will propagate into inaccuracies in the RSD power spectrum
$P^s(k_\parallel, k_\perp)$,  correlation function $\xi^s(r_\parallel,
r_\perp)$ and the hybrid statistics $P^s(k_\parallel, r_\perp)$.  For
brevity we only investigate its impact on $P^s(k_\parallel,
r_\perp)$. 

If the error $\Delta G$ has no imaginary part, and is independent of $r_{\parallel}$, it leads to
$\Delta P^s= P^s(k_\parallel=0,r_\perp) \Delta G=w_p(r_\perp)\Delta
G$. Since the absolute value of $\Delta G$ in the cumulant expansion
is in general $<0.01$ for $k<0.2\hmpc$, the resulting error in $P^s$
is $\la 1\%$. But the real situation is more complicated than that,
since $\Delta G$ is neither real nor independent of $r_\parallel$. For
this we have to numerically integrate over Eq. \ref{eqn:hybrid} to
obtain the resulting error in $P^s$. This integral involves the oscillating
integrand and is numerically challenging to reach better than $1\%$ in
$P^s$, making the accurate quantification of $\Delta P^s$
difficult. For this reason, in the current paper we only show the
error in the  integrand, induced by $\Delta G$. 

\begin{figure}
    \centering
    \includegraphics[width=12cm, angle=0]{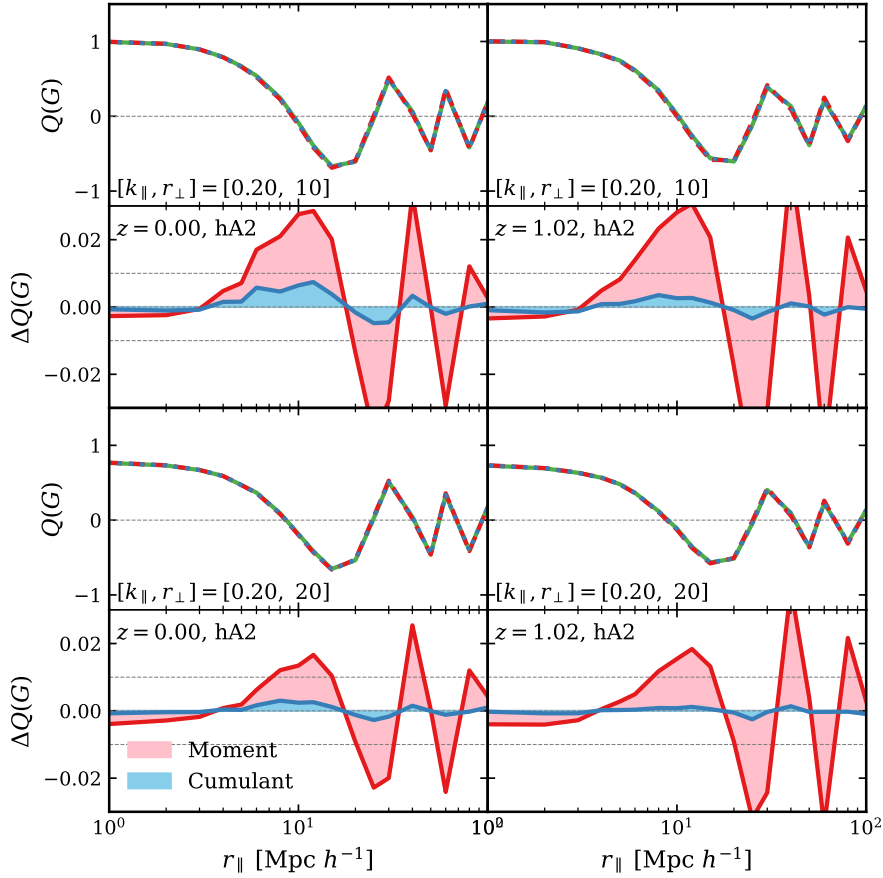}
    \caption{\textit{upper}: The integral kernel $Q(G)$ for halo set A2 at $z=0$ (\textit{l.h.s}) and $z=1$ (\textit{r.h.s}). \textit{lower}: residuals for the two different approaches.}
    \label{fig:Qg}
\end{figure}

Since  $P^s(k_{\|},r_{\bot})$ is real,
\ba
    P^s(k_{\|},r_{\bot})=\int Q(k_{\|},r_{\|},r_\bot)d r_{\|}\ .
\ea
Here the integrand
\ba
    Q&=&\left[(1+\xi(r))\mathrm{Re}G(k_{\|},r_{\|},r_\bot)-1\right]\cos(k_{\|}r_{\|})
    \nonumber\\
    &&-(1+\xi(r))\mathrm{Im}G(k_{\|},r_{\|},r_\bot)\sin(k_{\|}r_{\|})\ .
\ea
$Q$ in the simulation and the associated error $\Delta Q$ by the
moment/cumulant expansion to 4-th order are shown in
Fig. \ref{fig:Qg}. Since the largest error in $G$ occurs at
$r_\perp\la 10\mpch$, we only show the cases of
$r_\perp=10,20\mpch$. At $k_\parallel=0.2\hmpc$, $|\Delta Q|<0.01$ and
for most $r_\parallel$ $|\Delta Q|\ll 0.01$, for the cumulant
expansion up to the order of $\langle v_{12}^4\rangle_c$. For
comparison, we also show the case of moment expansion, whose error is
much larger. 

\subsection{Peculiar velocity decomposition}

\begin{figure}
    \centering
    \includegraphics[width=12cm, angle=0]{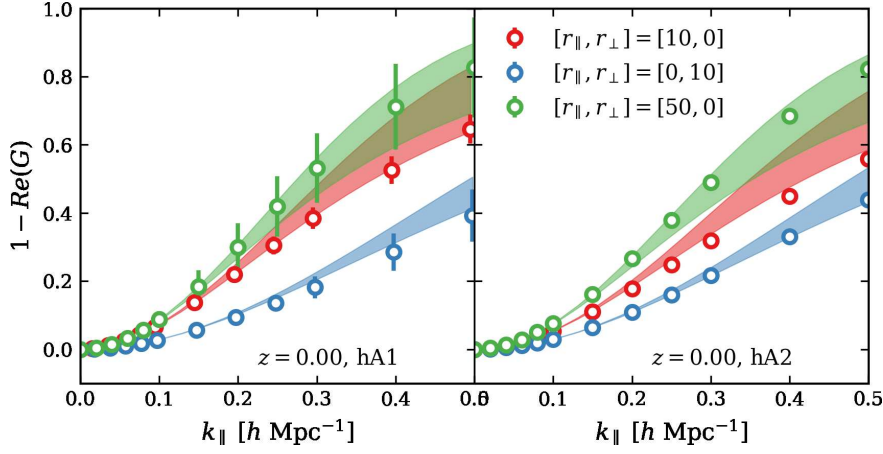}
    \caption{Test for Eq. \ref{eqn:gl_mgf}. The upper edge of the shaded region is Gaussian distribution limit ($\sigma_\mathcal{G}^2=\sigma^2$,$\sigma_\mathcal{E}^2=0$), the lower edge is exponential distribution limit($\sigma_\mathcal{E}^2=\sigma^2$,$\sigma_\mathcal{G}^2=0$).}
    \label{fig:lg_pdf}
\end{figure}

Fig. \ref{fig:lg_pdf} illustrates the results of Eq. \ref{eqn:gl_mgf} for halo sets A1 and A2 at $z=0$.
We first measure the velocity dispersion $\sigma_v$.
Then consider two extreme cases: Gaussian limit, $\sigma_\mathcal{G}=\sigma_v$ and exponential limit, $\sigma_\mathcal{E}=\sigma_v$.
The upper edge and lower edge of each shaded region correspond to Gaussian and exponential limit respectively.
The data points with error bars are direct measurements from halo catalogs.
At the non-linear regime (blue and red colored data in Fig. \ref{fig:lg_pdf}),
data points are close to the exponential limit.
Yet when move to the linear regime, as the green colored data shows, due to the scale is sufficiently large ($r_{\|}=50\mpch$) the results are close to the Gaussian limit.
The results suggest there is strong possibility that the pairwise velocity PDF is a mixture of Gaussian and exponential distributions.
At $k\sim 0.2\hmpc$ scale, it's no longer safe to take the Gaussian distribution assumption. 
This conclusion matches with above moment/cumulant expansion approaches.

\begin{figure}[!h]
    \centering
    \includegraphics[width=10cm, angle=0]{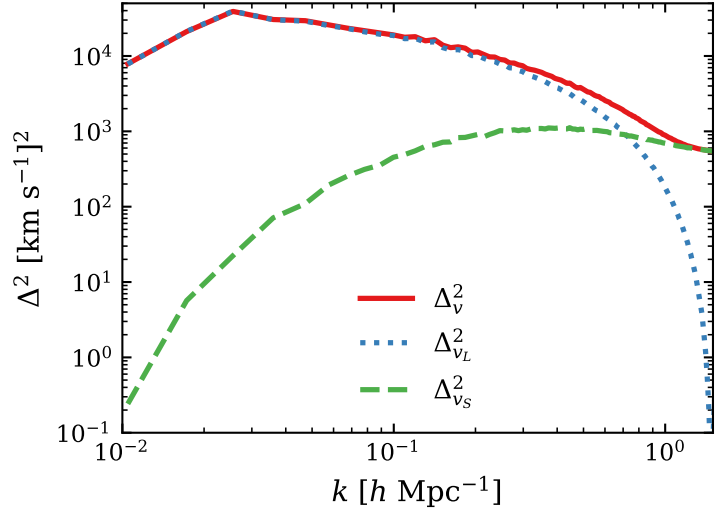}
    \caption{The dark matter power spectrum for peculiar velocity, deterministic component and stochastic component at $z=0$. $\Delta^2=k^3 P(k)/(2\pi^2)$.}
    \label{fig:vcomp}
\end{figure}

\begin{figure}[h!]
    \centering
    \includegraphics[width=14cm, angle=0]{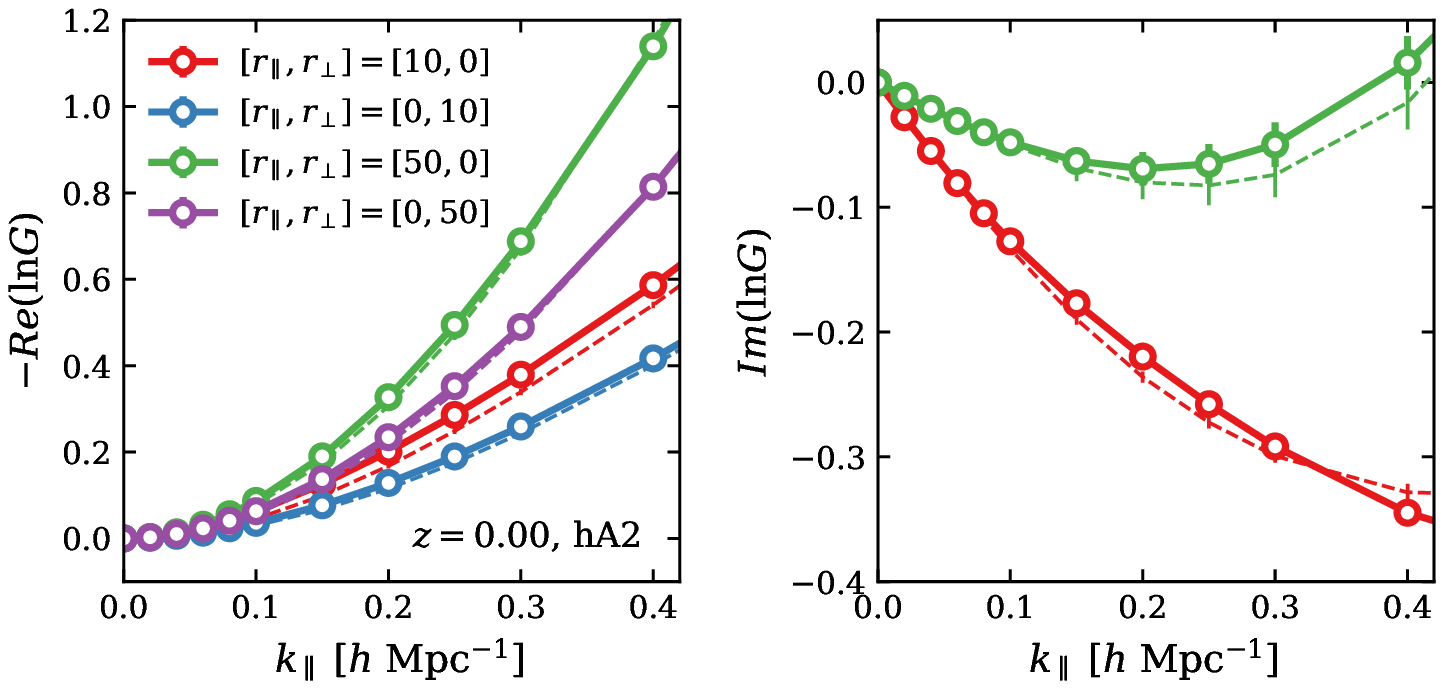}
    \caption{Test for Eq. \ref{eqn:ls_mgf}. Notice the vertical axis here is no longer about $G$ but $\ln{G}$. The solid lines with data points are $\ln{G^L}+\ln{G^S}$, where $G^L$ and $G^S$ is measured from simulation using the velocity decomposition method. The dashed lines are $\ln{G}$ directly measured from simulation.}
    \label{fig:ls_test}
\end{figure}

Fig.\ref{fig:vcomp} shows the dark matter power spectrum evaluated from N-body simulation for the total velocity field, as well as the two components introduced in \S 2.5 at redshift $z=0$.
The behavior of each component is consistent with the descriptions in \S 2.5. 
Fig. \ref{fig:ls_test} is the test for Eq. \ref{eqn:ls_mgf}. 
When $G^L$ and $G^S$ are fully independent, $\ln{G^L}+\ln{G^S}$(solid lines) should be strictly equal to $\ln G$(dashed lines).
The slight deviations implies there's correlation between the deterministic component and stochastic component, 
especially at the smaller scales (red and blue colored data).

\begin{figure}[h!]
    \centering
    \includegraphics[width=14cm, angle=0]{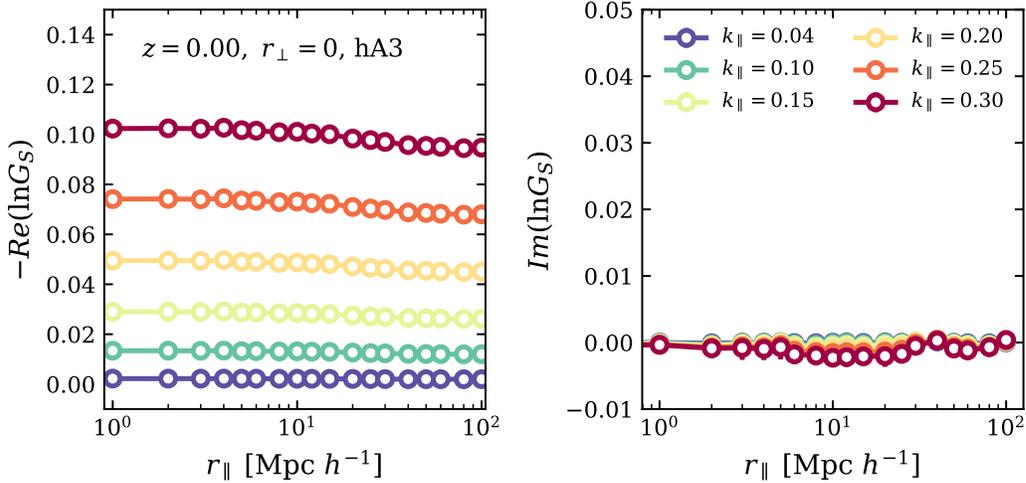}
    \caption{The moment generating function for stochastic component.}
    \label{fig:gstoc}
\end{figure}
Fig. \ref{fig:gstoc} shows the $\ln G^S$ as a function of $r_{\|}$ for halo set A3 at $z=0$ when fix $r_\bot=0$. 
We demonstrate A3 here because the random motion is more sever for the smaller halos, and therefore the stochastic component should be more significant than A1 and A2.
The results suggest the stochastic component is almost scale independent.
The real part Re$(\ln G_S)$ is decided by the $\langle v_{12,S}^m\rangle, m=2,4,6,\dots$. Since there is no cross-correlation between two different points 1 and 2 for a stochastic field, the scale dependent part in $\langle v_{12,S}^m\rangle$ vanishes, only the auto-correlation part resides. The non-zero value of Re$(\ln G_S)$ implies there is Gaussian component in stochastic velocity field.
For the imaginary part, as expected, it is not only scale independent but also zero.

\section{Conclusions and discussions}\label{sec:conclusion}
In this work, we investigate the convergence of measuring moment generating function in both Moment and Cumulant expansion approaches and find:
(1) Cumulant expansion performs much better than the Moment expansion for all halos samples and redshifts investigated.
(2) at $k< 0.1 \hmpc$ scale, including only the order of $n=1,2$ Cumulants is sufficient for modeling RSD.
(3) at $k\sim 0.2 \hmpc$ scale, the order of $n=1,2,3,4$ Cumulants must be considered. 
When considering the 3rd and 4th order pairwise velocity moments and cumulants, the cumulant expansion approach performs much better in the Hybrid statistics $P^s(k_\parallel,r_\bot)$.

Studies on the pairwise velocity PDF support a mixture of Gaussian and exponential pairwise velocity PDF. 
The results also support the above conclusions that the Gaussian streaming model only works at $k<0.1 \hmpc$.
RSD models based on $p(v_{12})$ can not take the Gaussian as well as exponential distribution assumptions at $k\sim 0.2 \hmpc$.
Further investigation on the peculiar velocity decomposition suggest a correlation between deterministic and stochastic components at small scale, and a Gaussian mixture part in stochastic component.

Comprehensive further investigations are required to implicate these findings in improving the RSD modeling.
In this work, by measuring $G$, 
we aim to investigate what is the requirement for the truncation of the peculiar velocity statistics in order to accurately model RSD, 
and the rationality and reliability to adopt Gaussian or exponential distribution assumptions to the pairwise velocity PDF.
Nevertheless, since the full understanding of the pairwise velocity PDF is still a long-standing problem in RSD cosmology, 
precisely building the association of the expansion coefficients for both moment expansion $\langle v_{12}^n\rangle$, and cumulant expansion $\langle v_{12}^n\rangle_c$, with the cosmological parameters are very difficult.
Moreover, in order to apply our method in practice, a complete calculation for the hybrid statistics $P^s(k_{\|},r_\bot)$ is required.
However, in this work, we mainly focus on the moment generating function $G$,
so we just briefly compare the errors of the integrand of Eq.\ref{eqn:hybrid} for the two different expansions.
Eq.\ref{eqn:hybrid} is an integral for an oscillatory function $Q$, 
the FFTLog method \citep{Hamilton2000} might be adopted when measuring $P^s(k_{\|},r_\bot)$.
We will have more detailed studies on the pairwise velocity PDF and $P^s(k_{\|},r_\bot)$ in our future works.

\begin{acknowledgements}
This work was funded by the National Natural Science Foundation of China (NSFC)
under No.11621303.
\end{acknowledgements}




\bibliographystyle{raa}
\label{lastpage}

\end{document}